\documentclass[11pt]{revtex4}
\usepackage{amsfonts}
\usepackage{amssymb,epsf}
\usepackage{latexsym}
\usepackage{epsfig}
\begin{document}

\title{Statistical pairing fluctuation and phase transition in $^{94}Mo$}
\author{Z.Kargar  \footnote{
kargar@susc.ac.ir }
 and V.Dehghani }
\address{Physics Department, College of Sciences, Shiraz University, Shiraz 71454,
         Iran\\}
\begin{abstract}
In the framework of BCS model, we have applied the isothermal probability distribution
to take into account the statistical fluctuations in calculation of
thermodynamical properties of nuclei. The energy and the heat capacity are calculated in $^{94}Mo$ nucleus using the mean gap parameter. The results
are compared with the values obtained based on the most probable values, experimental
data as well as some other theoretical models.
We have shown that heat capacity versus temperature behaves smoothly
instead of singular behavior predicted by
the standard BCS model. Also a smooth peak in heat capacity is observed which is a
signature of transition from normal to superfluid phase.
\end{abstract}

 \maketitle

\newpage
\section{Introduction}
Recently a lot of effort has been paid to describe the behavior of paired
small systems, such as metallic clusters and nuclei.
The starting point in all these methods is pairing hamiltonian of the system under study.
Various methods are used to calculate the partition(or grand partition) function
and they are different in how the constraint of fixed number of particles is considered in the calculations.
Some of the methods such as the Static-Path Approximation(SPA) plus Random-Phase Approximation(RPA) \cite{rr,kk} use number projected partition function.
Some others as Lipkin-Nogami method \cite{hj,yn,hc} add non-extensive terms to the free energy
to fix the particle number.
In the BCS method \cite{jb,lg2} grand partition function is used to obtain thermodynamic functions of the system. The BCS
uses the most probable value of pairing gap parameter which for few particle systems like nucleus leads to prediction
of some unreal singularities in heat capacity that is due to
of ignoring the impact of fluctuations in such systems. Experiments conducted by Oslo Group \cite{kke}
show a very smooth variation of heat capacity versus temperature, with a
peak which is believed to be due to a phase transition in nuclei.
To tackle the problem of singularities in BCS we must take into account the effect of
fluctuations. So we have used the isothermal probability distribution \cite{ld,lg}, which states the probability of
a system to be in a specific configuration is proportional to the exponential of it's free energy.
Using this principal we calculate the mean value of gap parameter and then the thermodynamic properties of nuclei based on
of the mean value of pairing gap parameter instead of the most probable value.

\section{Model}
The BCS is the standard theory to deal with a system of fermions having pairing potential.
The main tool which is used to obtain thermodynamic properties in this method is the
grand potential of the system, $\Omega$, which has
the following form in the mean field approximation:
\begin{equation}\label{omega}
\Omega= -\beta\sum(\varepsilon_k-\lambda-E_k)+2\sum\ln(1+\exp(-\beta E_k))-\beta\frac{\Delta^2}{G}
\end{equation}
where $\varepsilon_k$ is single particle energy of particles,$E_k=\sqrt{(\varepsilon_k-\lambda)^2+\Delta^2}$
is quasi particle energy, $G$ is the pairing strength, $\lambda$ is the chemical potential and $\beta=\frac{1}{T}$, where $T$ is the
temperature. In this equation $\Delta$ is the gap parameter which is a measure of the pairing correlation.
Thermodynamic quantities of the system, such as the number of particles $N$, the energy of the system $E$ and the entropy $S$ are calculated from following relations.
\begin{equation}\label{par}
N=\frac{\partial\Omega}{\partial\alpha}\ \ \ ,\ \ \ E=-\frac{\partial\Omega}{\partial\beta}\ \ \ ,\ \ \ S=\Omega-\alpha N+\beta E
\ \ \ \ \ \ \ \ (\alpha=\beta\lambda)
\end{equation}
The standard procedure to calculate gap parameter is minimizing the free energy
\begin{equation}\label{most}
\frac{\partial\Omega}{\partial\Delta}=0
\end{equation}
which leads to the BCS gap equation
\begin{equation}\label{gap}
\sum\frac{1}{E_k}\tanh(\frac{1}{2}\beta E_k)=\frac{2}{G}.
\end{equation}
Using this equation, the following relations for thermodynamic quantities will be obtained:
\begin{equation}\label{numb}
N=\sum[1-\frac{\varepsilon_k-\lambda}{E_k}\tanh(\frac{1}{2}\beta E_k)]
\end{equation}
\begin{equation}\label{en}
E=\sum\varepsilon_k[1-\frac{\varepsilon_k-\lambda}{E_k}\tanh(\frac{1}{2}\beta E_k)]-\frac{\Delta^2}{G}
\end{equation}
\begin{equation}\label{entr}
S=2\sum\ln[1+\exp(-\beta E_k)] +2\beta\sum\frac{E_k}{1+\exp(\beta E_k)}.
\end{equation}
The specific heat of the system, $C$, is obtained using the entropy relation. Neglecting the small change in $\lambda$ versus temperature, the specific heat will be
\begin{equation}\label{c}
C=\frac{1}{T}\frac{dS}{dT}=\frac{1}{2}\sum{sech^2(\frac{1}{2}\beta E_k)[\beta^2 E_k^2-\beta\Delta\frac{d\Delta}{dT}]}
\end{equation}
where
\begin{equation}\label{de}
\frac{d\Delta}{dT}=\frac{\frac{1}{2}\sum sech^2(\frac{1}{2}\beta E_k)}{\Delta(\frac{\beta}{2}\sum \frac{sech^2(\frac{1}{2}\beta E_k)}{E_k^2}-\sum \frac{\tanh(\frac{1}{2}\beta E_k)}{E_k^3})}.
\end{equation}
Applying the above formula for few body systems like nucleus dose not provide a good approximation of thermal properties since in calculation
of the gap parameter, the
most probable value of gap parameter is used while ignoring the important effect of thermal fluctuations on the behavior of
probability density function of $\Delta$, $P(\Delta)$.
Letting $P(\Delta)$ as the probability density of $\Delta$ to have a value
in temperature $T$, then according to Landau principal it takes the form
 $P(\Delta)\propto\exp(\Omega(\beta,\Delta))$
and the mean value of gap parameter will be
 \begin{equation}\label{me}
\bar{\Delta}=\frac{\int_0^\infty exp(\Omega(\beta,\Delta))\Delta d\Delta}{\int_0^\infty exp(\Omega(\beta,\Delta))d\Delta}
\end{equation}
Using $\bar{\Delta}$, the modified expressions for $E,N,S$ and $C$ will be
\begin{eqnarray}\label{nmea}
N=\frac{\partial\Omega}{\partial\alpha}&=&\sum[1-\frac{\varepsilon_k-\lambda}{E_k}\tanh(\frac{1}{2}\beta E_k)] \nonumber
\\
&+&\beta\bar{\Delta}\frac{\partial\bar{\Delta}}{\partial\alpha}(\sum\frac{\tanh(\frac{1}{2}\beta E_k)}{E_k}-\frac{2}{G})
\end{eqnarray}
\begin{eqnarray}\label{emea}
E=&-&\frac{\partial\Omega}{\partial\beta}=\sum\varepsilon_k[1-\frac{\varepsilon_k-\lambda}{E_k}\tanh(\frac{1}{2}\beta E_k)]-\frac{\bar{\Delta}^2}{G} \nonumber \\
&-&(\bar{\Delta}^2+\beta\bar{\Delta}\frac{\partial\bar{\Delta}}{\partial\beta})(\sum\frac{\tanh(\frac{1}{2}\beta E_k)}{E_k}-\frac{2}{G})
\end{eqnarray}
\begin{eqnarray}\label{smea}
S&=&2\sum \ln [1+\exp(-\beta E_k)]+2\sum \frac{\beta E_k}{1+\exp(\beta E_k)} \nonumber \\
&+&\beta^2\bar{\Delta}(\lambda\frac{\partial\bar{\Delta}}{\partial\alpha}+\frac{\partial\bar{\Delta}}{\partial\beta})
(\frac{2}{G}-\sum\frac{\tanh(\frac{1}{2}\beta E_k)}{E_k})
\end{eqnarray}
where $E_k=\sqrt{(\varepsilon_k-\lambda)^2+\bar{\Delta}^2}$. The last term in the above equations is the product of gap equation by other
 terms, which  is absent in the standard BCS formulation. According to this procedure the heat capacity will be
\begin{eqnarray}\label{cmea}
C=&-&\beta\frac{dS}{d\beta}
=\frac{1}{2}\sum{sech^2(\frac{1}{2}\beta E_k)[\beta^2 E_k^2+\beta^3\bar{\Delta}\frac{d\bar{\Delta}}{d\beta}]}-
\beta(2\beta\bar{\Delta}\lambda\frac{\partial\bar{\Delta}}{\partial\alpha}+\beta^2\lambda\frac{d\bar{\Delta}}{d\beta}\frac{\partial\bar{\Delta}}{\partial\alpha} \nonumber \\
&+&\lambda\beta^2\bar{\Delta}\frac{d}{d\beta}\frac{\partial\bar{\Delta}}{\partial\alpha}
+2\beta\bar{\Delta}\frac{\partial\bar{\Delta}}{\partial\beta}+
\beta^2\frac{d\bar{\Delta}}{d\beta}\frac{\partial\bar{\Delta}}{\partial\beta}+\beta^2\bar{\Delta}\frac{d}{d\beta}\frac{\partial\bar{\Delta}}{\partial\beta})
(\frac{2}{G}-\sum\frac{\tanh(\frac{1}{2}\beta E_k)}{E_k}) \nonumber \\
&+&\beta^3\bar{\Delta}(\lambda\frac{\partial\bar{\Delta}}{\partial\alpha}+\frac{\partial\bar{\Delta}}{\partial\beta})\sum[\frac{sech^2(\frac{1}{2}\beta E_k)}{E_k}(\frac{E_k}{2}+\frac{\beta}{2}\frac{\bar{\Delta}}{E_k}\frac{d\bar{\Delta}}{d\beta})-\frac{\bar{\Delta}}{E_k^3}\frac{d\bar{\Delta}}{d\beta}\tanh(\frac{1}{2}\beta E_k )].
\end{eqnarray}

\section{Results And Discussion}
In this work we assume neutrons and protons as
two distinct non-interacting thermodynamic systems. Single particle energies of
Nilsson potential with a specified quadrapole deformation, are used.
\begin{figure}[th]
\begin{center}
\begin{picture}(210,100)
\put(0,-40){ \epsfxsize=7cm \epsfbox{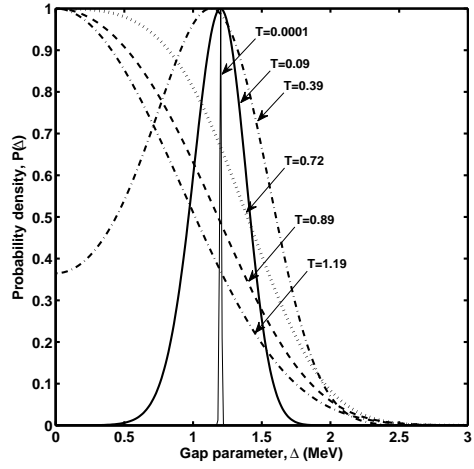} }
\end{picture}
\end{center}
\vspace*{1cm} \caption{Dependence of the probability density function $P(\Delta)$ on the gap parameter $\Delta$ for various temperatures
for neutrons in $^{94}Mo$ nucleus.}\label{F1}
\end{figure} We
have taken the deformation parameter to be, $\beta=0.15$ for $^{94}Mo$ nucleus \cite{kke} and we have used the following parameterizations
as are used in the model for $k$ and $\mu$ \cite{sg}, the spin-orbit and centrifugal parameters
\begin{eqnarray}\label{kmu}
k_n= 0.06385 &;& \mu_n= 0.508004 \\
k_p= 0.06928 &;& \mu_p= 0.554006 .
\end{eqnarray}
\begin{figure}
\centering
\begin{tabular}{cc}
\epsfig{file=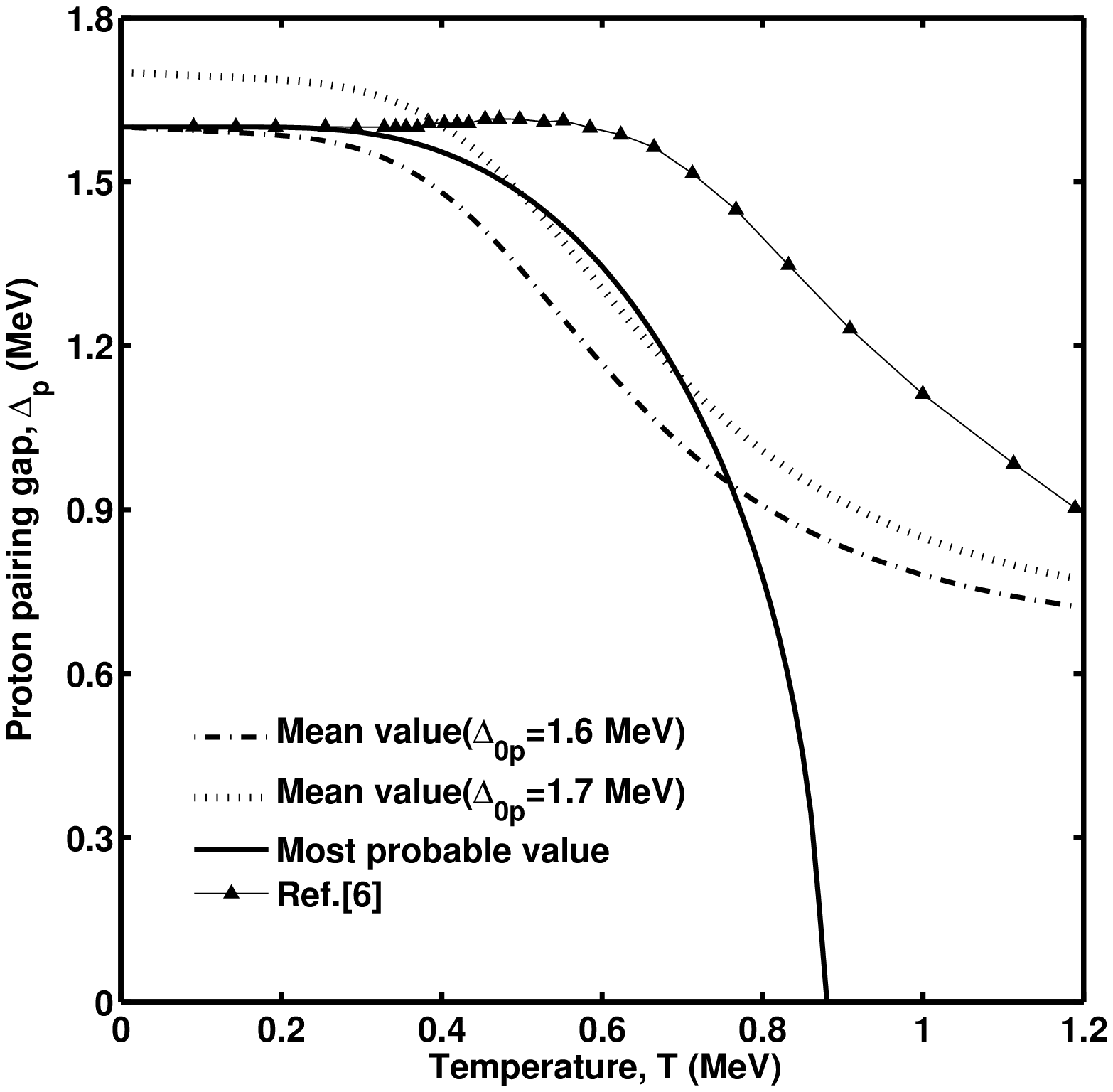,width=0.4\linewidth,clip=} &
\epsfig{file=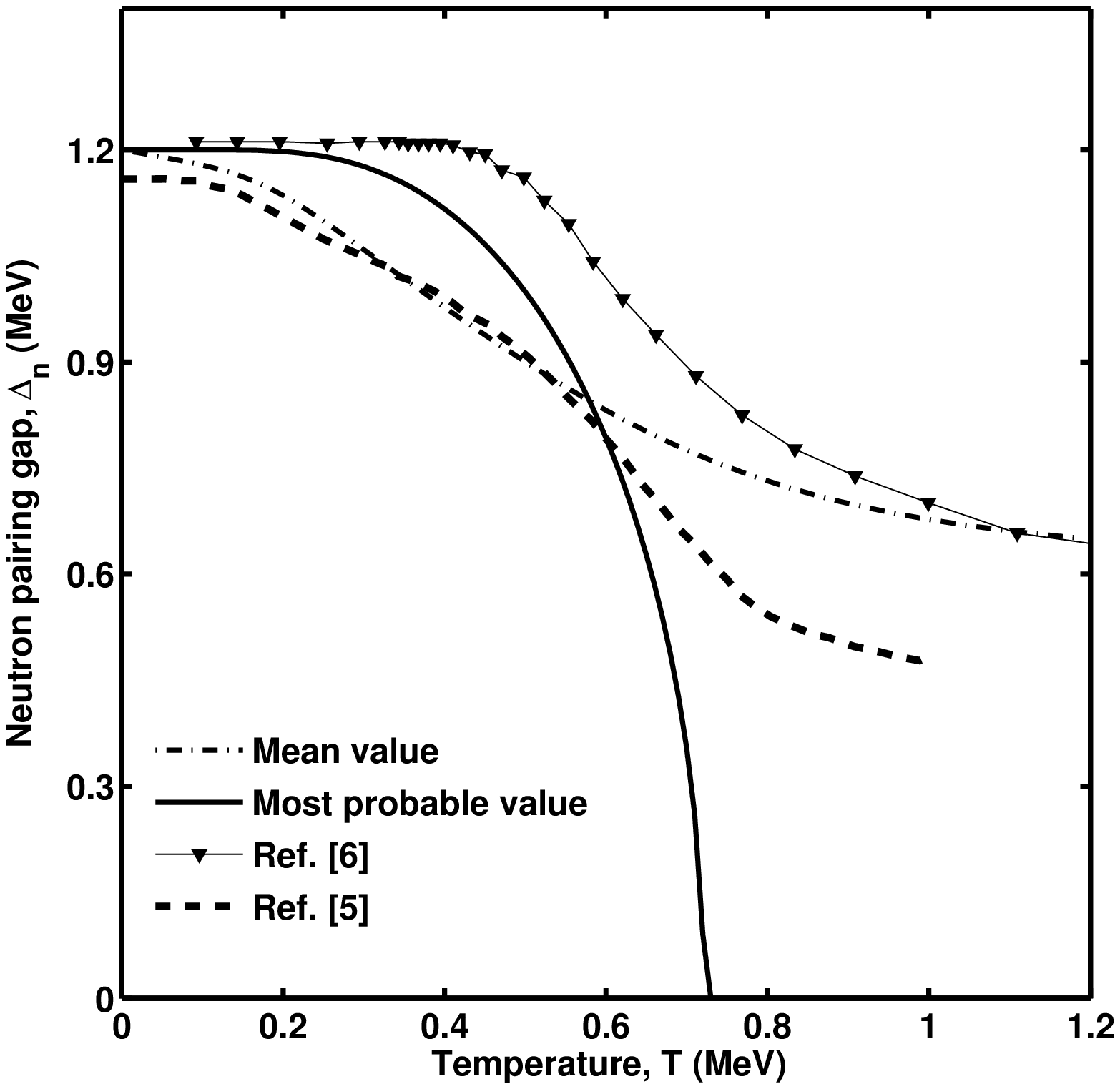,width=0.4\linewidth,clip=} \\
\end{tabular}
\vspace*{0cm} \caption{Temperature dependence of gap parameter in $^{94}Mo$ nucleus for the proton and the neutron system.}\label{F2}
\end{figure}
The values of the gap parameter at $T=0$ are $\Delta_{0p}=1.6 Mev$ and $\Delta_{0n}=1.2 Mev$ \cite{kke}.
which are used to calculate the interaction strength, $G$, by solving the gap equation (\ref{gap}) and equation (\ref{numb}) for protons and neutrons at zero
temperature. In order to calculate $P(\Delta)$ as a function of temperature for
both types of particles, particle number constraint, equation (\ref{nmea}) is used.
The results for neutrons in $^{94}Mo$ nucleus are given in fig.(\ref{F1}) in which $P(\Delta)$ is normalized.
As it is seen its shape changes from symmetric Gaussian at
temperatures lower than critical temperature $T_c=0.72 MeV$, the temperature in which gap parameter becomes zero, We have extracted the following values of pairing gap at zero temperature, $\Delta_0$ for
protons and neutrons, using the three point method \cite{pm} to asymmetric shape at temperatures higher than $T_c$.
In all temperatures the maximum value of $P(\Delta)$ is consistent with that of the most probable values.
The resulted values of $\Delta$ versus temperature for protons and neutrons in $^{94}Mo$ nucleus are plotted in fig.(\ref{F2}).
We have extracted the following values of pairing gap at zero temperature, $\Delta_0$ for
protons and neutrons, using the three point method \cite{pm}
\begin{eqnarray}\label{kmude2}
\Delta_{0p}=1.7Mev &;& \Delta_{0n}= 1.15Mev
\end{eqnarray}
The results are also shown in fig.(\ref{F2}). The effect of different values of $\Delta_0$ on heat capacity will be discussed. The gradual decreasing of gap parameter with a sudden decrease is seen which can be comparable with the experimental values \cite{kk}. The results obtained by K. Kaneko et.al,\cite{kke}, are also shown for comparison.
This is interpreted as a rapid breaking of nucleon Cooper pairs and the suppression of pairing correlation. Also this is correlated with
the S shape of the heat capacity.
\begin{figure}
\begin{center}
\begin{picture}(210,100)
\put(0,-40){ \epsfxsize=7cm \epsfbox{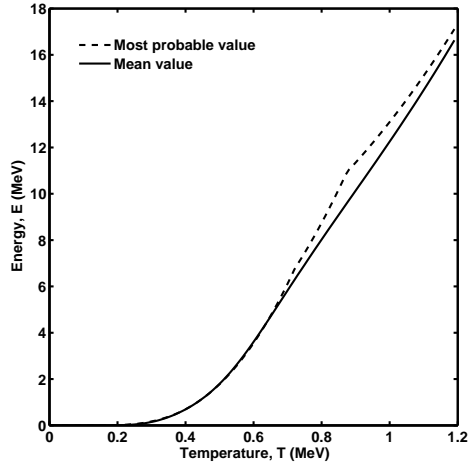} }
\end{picture}
\end{center}
\vspace*{1cm} \caption{Total energy versus temperature using the most and mean probable gap parameter in $^{94}Mo$ nucleus.}\label{F3}
\end{figure}
The mean gap parameter $\bar{\Delta}$ and its partial derivatives were applied to calculate energy as a function of temperature using equation(\ref{emea}) the total energy is $E=E_n+E_p$.
The results are plotted in fig.(\ref{F3}) in comparison with the most probable energy from equation(\ref{en}). This figure shows how using
the mean value of the gap parameter leads to the smooth behavior of the energy near the critical temperature.

\begin{figure}[th]
\begin{center}
\begin{picture}(210,150)
\put(0,-40){ \epsfxsize=7cm \epsfbox{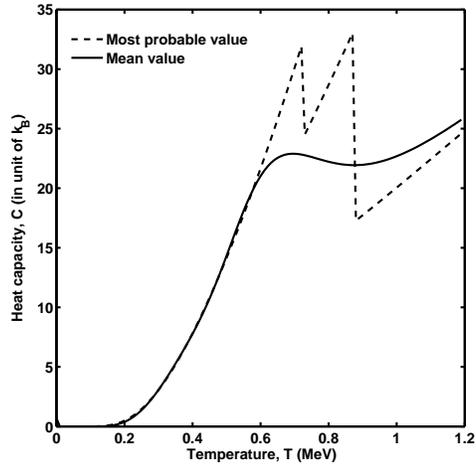} }
\end{picture}
\end{center}
\vspace*{1cm} \caption{Heat capacity as a function of temperature using the most and the mean probable gap parameter in $^{94}Mo$ nucleus.}\label{F4}
\end{figure}
\begin{figure}[th]
\begin{center}
\begin{picture}(210,150)
\put(0,-40){ \epsfxsize=7cm \epsfbox{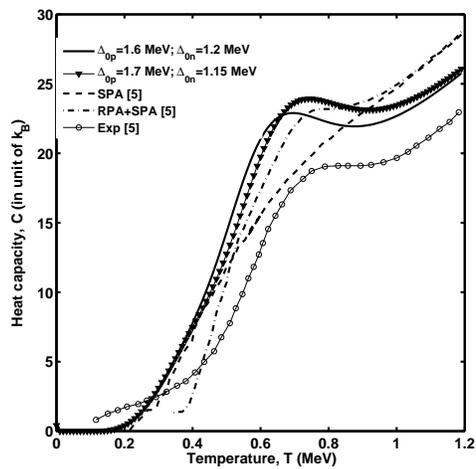} }
\end{picture}
\end{center}
\vspace*{1cm} \caption{Heat capacity versus temperature using the mean probable gap parameter in $^{94}Mo$ nucleus. The experimental values and the calculated
values from SPA and RPA+SPA methods(dashed and dotted-dashed lines, respectively, taken from \cite{kk}) are also plotted for comparison.}\label{F5}
\end{figure}
Based on the mean gap parameter, the heat capacity is calculated using equation (\ref{cmea}). The results are plotted
in fig.(\ref{F4}), for $^{94}Mo$ nucleus, for the two sets of $\Delta_0$, where $C=C_n+C_p$. Also in fig.(\ref{F5}) the experimental \cite{kke} and the RPA+SPA heat capacities \cite{kk} are
plotted for comparison.
The extracted results show the discontinuities have been disappeared and the heat capacity exhibits an S shape, very similar in shape with the
experiments above $0.4 MeV$. Below $0.5 MeV$, the results of SPA, the NPSPA and this work are in coincidence.

In summery, taking into
account thermal fluctuation, the discontinuity of the heat capacity is washed out and a continues S-shape around the critical temperature has been obtained
as it is found experimentally \cite{kke}. The S shape of the heat capacity is well correlated in temperature with the suppression of $\Delta_p$ and $\Delta_n$
as is shown in fig.(\ref{F2}). This is interpreted as a signature of the thermal pairing phase transition.

\end{document}